\documentclass[12pt,preprint]{aastex}

\shorttitle{The  dynamical  simulations  of  the  planets orbiting
GJ 876} \shortauthors{Ji Jianghui et al.}

\received{2001 December 5}
\begin{document}

\title{The  dynamical  simulations  of  the  planets orbiting  GJ 876}

\author{JI  Jianghui\altaffilmark{1,3} and LI  Guangyu\altaffilmark{1,3}}
\affil{Purple  Mountain  Observatory , Chinese  Academy  of
Sciences,Nanjing,210008}
\email{jijh@pmo.ac.cn}

\and

\author{LIU  Lin\altaffilmark{2,3}}
\affil{Department of Astronomy, Nanjing University, Nanjing ,210093}

\altaffiltext{1}{Purple  Mountain  Observatory , Chinese  Academy  of  Sciences
,  Nanjing , 210008,  China}
\altaffiltext{2}{Department of Astronomy,  Nanjing University,  Nanjing ,
210093, China}
\altaffiltext{3}{National  Astronomical  Observatory ,  Chinese  Academy  of
Sciences,China}

\begin{abstract}
  In this paper we have performed simulations to investigate the dynamics of the
M dwarf star GJ 876 in an attempt to reveal any stabilizing mechanism for
sustaining the system. We simulated different coplanar and non-coplanar
configurations of  two-planet system and other cases. From the simulations,
we found that the 2:1 mean motion resonance between  two planets
can act as an effective mechanism of maintaining the stability of the system.
This result is explained by a proposed analytical model.
By means of the model, we still studied the region of  motion of the inner
planet by varying the parameters of  the system and detected that the analytical
results are well consistent with the numerical simulations.
\end{abstract}

\keywords{dynamical simulations,stellar dynamics -methods:N-body simulations,
planetary systems-stars:individual GJ 876
}

\section{Introduction}

 At  present more and more extrasolar planetary systems are being discovered
(Marcy et al. 2000; Butler et al. 2000), and many research groups throughout the world
are now devoting themselves to surveying these planets.
From  the  Extrasolar  Planets Encyclopaedia website(http://www.obspm.fr/planets),
we have known  that $\sim$60 planets around main-sequence stars are confirmed
and  there are many others yet to be confirmed.  Most  of  the  extrasolar
planets are found  to either orbit very close to their host stars or  travel on
much more eccentric  paths  than  any of  the major planets in our  solar
system, which is an enormous challenge for standard  planetary  formation
theories as well as for the dynamical stability  of  the  planetary  system.

Marcy et al. (2001) pointed out that the two planets orbiting the
M4 main-sequence star GJ 876 are now apparently locked in the
state of 2:1 resonance, with orbital periods 30.1\,d and 61.0\,d,
and semimajor axes 0.13\,AU\ and 0.21\,AU\ . They suggested that
stability might be sustained by the above mean-motion resonance.
In their paper,they  gave GJ 876 an estimated mass of $0.32 \pm
0.05M_{\odot}$, and this value was used in our numerical
simulations. The masses of the two planets given in Table 1
(Laughlin \&  Chambers 2001) were derived with an inclination
sin\textit{i} = 0.55, where \textit{i} is  the inclination of  the
orbit relative to sky plane.The masses of the planets are always
utilized except for special comments.

Laughlin \&  Chambers (2001) showed  that short-term perturbations
among massive planets in multiple planet system can give rise to
radial  velocity  variations of  the  central  star that  differ
from  those  which  consider  the  planets  move  as  Keplerian
ellipse. They also  implied  that  the  configurations  given in
the paper  are  stable for  at  least  $10^7$  yr  as  the mean
motion resonances  and  secular resonances  appeared during the
long-term orbital evolution. Recently  Rivera \& Lissauer (2000)
studied the planetary system orbiting $\upsilon$ Andromedae,which
consists of three Jovian-mass planets with  the orbital  periods
range from  $\sim$4 d to $\sim$4 yr. They found  that some
configurations are stable for at least $10^9$ yr, but others can
be ejected into the stellarspace because of the excitation of  the
eccentricity. Due to their great efforts, now  we have more
helpful knowledge about the dynamics of the extrasolar systems,
however in this article we shall present our substantial research
of  GJ 876 by simulations.

 In  the rest  of  this  paper is  organized as follows:  In Section 2 ,we
briefly introduce the numerical setup of the dynamical simulations. In Section
3, we present the results of  the simulations of  the GJ 876 system. In Section 4,
we describe  an analytical model that helps to explain the resonant mechanism of
the studied system. In Section 5 , we give a  brief discussion.

\section{Numerical setup}

We have developed a purely gravitational N-body scheme for
numerical simulation, in which general relativistic effect of the
central body is not considered at present. We adopted
RKF7(8)(Fehlberg 1968) to carry out the integrations. We used time
step of 0.3d ($\sim $ one percent of the orbital period of the
inner planet) when integrating two planets in our simulations. The
integrator was also optimized for close encounters during the
orbital evolution. The integration was automatically ceased if
either of the planets is deemed to be too far from the central
star. We also effectively controlled the numerical errors,with the
local truncation error $10^{-14}$, over our time span of
integration of 10$^{6}$ yr( $\sim $10$^{7}$ orbital period of the
inner planet). Additionally,as the energy of the system is
dissipative by using traditional algorithm ,so we again used
symplectic algorithm(Feng 1986;Wisdom \& Holman 1991),which has
many advantages such as holding the symplectic structure of the
Hamiltonian system and remaining the periodical variations of the
error of the energy, to examine several examples and then to
confirm the results given by RKF7(8). Therefore,on the basis of
these, we began to prepare the simulations of the system under
study.

On the whole,we carried out three groups of the simulation. In the
first runs, we aim to explore the likely resonance of the
planetary system. The two planets are considered to locate at the
same orbital plane, and we present herein one set of the
parameters of numerical simulations of planetary configurations
which are partly taken from Laughlin \& Chambers(2001) (see Table
1). It is clear that six orbital elements are respectively
semimajor axis $a$ , eccentricity $e$ , inclination $i$, nodal
longitude $\Omega $, periastron $\omega $ and mean anomaly $M$.
However, in the table the inclinations of the planets referred to
reference plane are constructed as well as the nodal longitudes.
We note that Fig. 1 shows the initial configuration of the two
companions. Moreover,in the simulations,we fixed the semimajor
axes, eccentricities and inclinations of the two planets and
developed codes to randomly generate other three angles of each
planet to furnish initial elements for integration.

For the second runs, we integrated the mutual inclined orbits to
investigate the dynamical characteristic of the system. In the
simulations, the cloned orbits of the outer planet are generated
by changing its inclination. Finally, we examined the cases that
the system is assumed to be composed of three planets. Rivera \&
Lissauer (2000) investigated the simulations of the test bodies of
$\upsilon$  Andromedae system. By comparison, our codes differ
from theirs in that we take the assumed planet as a massive body
that travels outside the orbit of the outer planet, which would be
expected to provide certain information to detect unknown planets
of the extrasolar planetary systems.

\section{Results  of  dynamical simulations}
 In this section, we will present the leading results of our dynamical simulations.
Our goal is to explore for different planetary configurations, the long-term
orbital evolution of GJ 876 and to attempt to reveal any mechanism that
helps to maintain the stability of the planetary system. Although
simulations may differ from reality, nevertheless they can be supposed to
exhibit some rough picture of the system under study.

\subsection{Two-planet  coplanar  system}
As is well known, most of the major planets in the solar system
have small inclinations with regard to the essential reference
plane. Thus at first we are very interested in the simulations of
the coplanar cases of the system. We performed an ensemble of
systems by combining the angles of $\Omega$, $\omega$ and $M$,
which the two planets have small inclinations, but the semimajor
axes, eccentricities and inclinations still remained unchanged for
all cases when we started to separately integrate these systems in
which the time span covers 1 Myr. In particular, Table 1 are
listed one set of the initial orbital parameters selected from
many examples.

By analyzing the integration results, we found that nearly 1/4 of
hundreds of systems in our simulations remain stable over a time
span of 1Myr. It is not difficult to understand that the stability
of a system is sensitive to its initial planetary configuration.
The results of the simulations show that there is a tendency for
many systems to self-destruct in 10$^{2}$-10$^{3}$ yr, and the
lifetime is even shorter for those unstable cases. Marcy et al.
(2001)  tried the similar simulations with different values of the
initial epoch of the integration to point out that the initial
epoch for the integration is an important factor in determining
the system stability. However, we are particularly concerned with
the stable cases. Table 2 lists the computational results by using
the initial parameters produced by Table 1. In the table are
presented the variations of the semimajor axis \textit{a}
,eccentricity \textit{e} and inclination \textit{i} of the two
planets for 1 Myr, one can observe that these orbital parameters
do not change dramatically and just remain bounded instead. In
addition, Fig. 2 displays the orbital variations of the inner and
outer planets for a stable case; we particularly see that the
semimajor axis, eccentricity and inclination undergo small
oscillations for the whole time span.We should point out that the
evolution shown in Figure 2 and the results of Table 2 are typical
of all the stable cases.

 The stabilizing mechanism of the system is really our interest, so we will
next focus on the resonant mechanism. In the usual notation of celestial
mechanics, the critical argument $\sigma$ for the mean motion resonance is

\begin{equation}
\label{eq1}
\quad\quad\sigma = \lambda _{1} - 2\lambda _{2} + \tilde {\omega} _{1},
\end{equation}

\noindent where
\begin{equation}
\label{eq2}
\quad\quad\lambda _{1} = \omega _{1} + \Omega _{1} + M_{1},
\end{equation}
\begin{equation}
\label{eq3}
\quad\quad\lambda _{2} = \omega _{2} + \Omega _{2} + M_{2},
\end{equation}
\begin{equation}
\label{eq4}
\tilde{\omega} _{1} = \omega _{1} + \Omega _{1}.
\end{equation}
\noindent
Here $\lambda _{1} $, $\lambda _{2} $ are , respectively, the mean
longitudes of the inner and outer planets, $\tilde {\omega} _{1} $ denotes
the longitudes of the periastron of the inner(hereafter subscript 1 denotes
the inner planet, 2 the outer planet). For the whole integration time span,
Fig. 3 exhibits the critical argument $\sigma$ librates with an amplitude of
$ \pm 70^{ \circ} $. It is easy to understand that because of the libration
of the critical argument, when the two planets are in conjunction, they are
far from each other and hence are protected from close encounters, resulting
in their mutual perturbations being weaker than in the non-resonant case.

   Again,we illustrate the motion of the inner planet in phase space. The
semimajor axis (action) of the inner planet is shown as a function of the
resonant argument $\sigma $ in Fig. 4. The action-angle coupling is
characteristic of the so-called ideal resonance. Notice that the equilibrium
point appears to be (0.130, 0), and this confirms the fitting result given
by Marcy et al.(2001)(ref. Table 3 in Marcy et al. 2001).Besides, the figure
reveals the fact that the 2:1 mean motion resonance between  two planets
corresponds to a moderate resonance in which the semimajor axis is well locked,
and further indicates that the existing resonance is an effective mechanism
to maintain the stability of the system. For the sake of better understanding,
we shall construct an analytical model to explain this in the next section.

  Meanwhile, for two-planet coplanar case, we carried out an extended,
forward integration to check the dynamical evolution of the system. The
result again suggested the 2:1 mean motion resonance. We also integrated
some other cases in order to examine whether the system could be stable when
the masses of the planets changed simultaneously. The amplitude in the
variation in radial velocity of the star (Laughlin \& Chambers 2001)
suggested minimum masses of $0.56M_{J} $ for the inner planet and $1.89M_{J}
$ for the outer one; we used the coupled values for the masses and the
orbital elements still remained unchanged to repeat the integration. The
results again showed that the 2:1 mean motion resonance still exists even
when the initial masses were varied. Therefore, it seems that this resonance
is possibly the most important dynamical feature of the system.

  However, one may ask whether there exist other likely mechanisms to sustain
the system, the answer is that the probability indeed could not be
ignored. The dynamics of the asteroids in our solar
system(Morbidelli \& Nesvorny 1999 ; Malhotra et al.
1996,1998,2000; Duncan \& Levison 1997) shows that the secular
resonances with the major planets(such as Jupiter, Saturn,
Neptune, etc) are very important for the long-term orbital
evolution of the asteroids, which can excite their eccentricities
and inclinations over tens of millions years. The leading
viewpoints are that the asteroids in the main belt can be
transported to near-earth space and the bodies of the Kuiper Belt
can be as a reservoir of the jovian short-period comets through
the complicated interaction of the mean motion resonance and
secular resonances. In addition, there is a special secular
resonance, so-called Kozai resonance(Kozai 1962), which
corresponds to the coupled oscillations of the inclination and
eccentricity of the planet, unlike the precession of the
longitudes of the bodies' perihelia(or periastra) or ascending
nodes. The discovery of the extrasolar planets provide abundant
sources to investigate the dynamics of the bodies. Laughlin \&
Chambers (2001) performed dynamical simulations to suggestion a
secular resonance of the GJ 876 system. Hence, we were determined
to focus on these secular resonances ,as a result , we were
capable of discovering such systems from the simulations. Fig.5
displays that the periastron of the inner planet temporarily
librates around 0$^{0}$ or 180$^{0}$, correspondingly, the
eccentricity is extremely excited to pump up to the value of
unity,which is associated with Kozai resonance. Additionally, the
diagram implies that this kind of secular resonance plays
significant part in the long-term dynamical evolution of the
planet.

\subsection{Two-planet highly inclined orbits system }
For other simulations, we again performed tests of highly inclined orbits
configuration to examine the dynamics of the GJ 876 system. Regarding our
solar system, we recollect the fact that most of the major planets have low
inclinations and eccentricities, accompanying the sun, therefore in a
viewpoint of dynamics the investigations of the non-coplanar cases can
further help understanding the dynamical evolution of the extrasolar
systems.

For the sake of simplicity , we only varied the inclination of the outer
planet but remained that of the inner planet. For each case, the inclination
of the outer planet was separately increased by 5$^{0}$, 10$^{0}$, 15$^{0}$,
20$^{0}$, 25$^{0}$ and 30$^{0}$ with respect to that of the inner planet.
Additionally, in these experiments, we only adopted other orbital elements given
by Table 1 rather than combined the angles. The numerical integrations
reveal the fact that for some cases, though the inclination was greatly
changed but the two planets are still in the mean motion resonance(MMR) ,
which indicates that to some extent this resonant mechanism can make the
system stable, while for unstable cases we found that the inner planet
abruptly leaves the system on account of the accumulation of perturbations by
the outer planet. Table 3 reports the details. However, these simulations
partly show some of dynamical features of the system, we can not simply
believe that highly inclined orbits will lead to the unstable systems by
common judgement, while the stability of the system is still relative to the
given initial conditions. Our conclusions are consistent with the dynamical
simulations of Marcy et al. (2001), who announced the finding that the two
planets were locked in a 2:1 mean motion resonance for all the stable
configurations.


Still,we also considered the cases of  three-planet system,and
details of the problem will be examined in a later paper(Ji \&
Liu, in preparation).

\section{Analytical model}
In Section 3, we have taken into account several groups of the planetary
configurations by numerical simulations , then we are strongly impressed that
the above-mentioned mean motion resonance between two companions could play important
role in the stabilization of the system, then it is essential for one to
clarify the mechanism of the problem. Hence, we propose an analytical model
to explain the mechanism of the mean motion resonance of the explored
system.

\subsection{Ideal resonant model of the problem}
From Table 1, we underline that these two planets have very low inclinations with
respect to the fundamental plane and the outer planet  has a small
eccentricity, and this reminded us of the ideal resonant model (Garfinkel
1966; Liu et al. 1985a,1985b) under the dynamical framework of a
star-plus-two planets system. For simplicity, if we take the mass of the
central star $M_{c} $ as the unit of mass, the semimajor axis of the outer
planet $a_{2} $ as the unit of length, and define the unit of time by
$\left[ {T} \right] = \left( {{{a_{2}^{3}}  \mathord{\left/
{\vphantom {{a_{2}^{3}}  {GM_{c}} }} \right. \kern-\nulldelimiterspace}
{GM_{c}} }} \right)^{{{1} \mathord{\left/ {\vphantom {{1} {2}}} \right.
\kern-\nulldelimiterspace} {2}}}$, then the gravitational constant $G$
will have the value unity.

In the beginning, we introduce Delaunay variables commonly used in the
celestial mechanics ,

\begin{equation}
\label{eq5}
\left\{
\begin{array}{lcl}
L = \sqrt {a}  \quad &,& \quad  l = M  \quad , \\
G = L\sqrt {1 - e^{2}} \quad  &,&  \quad g = \omega \quad ,  \\
H = G \cos i  \quad &,& \quad h = \Omega \quad . \\
\end{array}
\right.
\end{equation}

In the next step, we conveniently introduce the following transformation,

\begin{equation}
\label{eq6}
\left\{
\begin{array}{lcl}
\tilde {L} = L/q \quad & , & \quad \tilde{l} = ql - pl_{2}  +  p\left( {\tilde {\omega}  - \tilde {\omega} _{2}}  \right) \quad ,  \\
\tilde {G} = G - p\tilde {L}  \quad &,&  \quad \tilde {g} = g \quad , \\
\tilde {H} = H - p\tilde {L} \quad &,&  \quad \tilde {h} = h\quad . \\
\end{array}
\right.
\end{equation}

\noindent
where $p$ and $q$ are integers, and we let $q:p = 1:2$ in our paper.

Following our earlier paper (Ji, Liu \& Liao 2000), we will have the
action-angle variables(hereafter subscript 1 is omitted for simplicity):

\begin{equation}
\label{eq7}
\left\{
\begin{array}{l}
 \tilde {L} = \sqrt{a}  \\
 \tilde {l} = \sigma = \lambda-2\lambda_{2} + \tilde{\omega}  \\
\end{array}
\right.
\end{equation}

\noindent
Then in terms of the action-angle variables of Eqs. (\ref{eq6}) and (\ref{eq7}),
the relevant Hamiltonian function can be written as
\begin{equation}
\label{eq8}
 F = \tilde {F}_{0} \left( {\tilde {L}} \right) + \tilde {F}_{c} \left(
{\tilde {L}, \tilde {G}} \right) + F_{2/1} \cos\tilde {l}
\end{equation}

\noindent
where

\begin{equation}
\label{eq9}
\left\{
\begin{array}{l}
\tilde{F}_{0} ({\tilde{L}})\quad  = \quad \frac{{1}}{{2\tilde {L}^{2}}} + 2n_{2} \tilde {L}   \\
\tilde {F}_{c} ( {\tilde {L},\tilde {G}}) = \quad \mu _{2} a^{2}\left[{\frac{{1}}{{4}}\left( {1 + \frac{{3}}{{2}}e^{2}} \right) + \frac{{9}}{{64}}a^{2}\left( {1 + 5e^{2}} \right)} \right] \quad ,  \\
\tilde {F}_{1}  = F_{2/1} = \quad - \mu _{2} a^{2}\left( {\frac{{9}}{{4}} + \frac{{5}}{{4}}a^{2}} \right)e \quad , \\
\tilde {L} =  L  \quad \quad , \quad \tilde{G} =  G - 2\tilde {L}\quad .  \\
\end{array}
\right.
\end{equation}

\noindent where in Eq. 9, $\mu _{2} $ and $n_{2} $ ,respectively,
are the mass and mean motion rate of the outer planet.We should
point out that short period terms have been averaged, and that
high eccentricity terms are neglected when deriving Eqs. 8 and 9.

 As mentioned in the earlier papers(Liu et al. 1985a,1985b;Ji, Liu \& Liao
2000), the Hamiltonian $F$ does not explicitly involve $\tilde {g},\quad
\tilde {h}$ and $t$, therefore there exist the following integrals:

\begin{equation}
\label{eq10}
\tilde {G} = \tilde {G}_{0} ,
\quad
\tilde {H} = \tilde {H}_{0} ,
\end{equation}

\noindent
and

\begin{equation}
\label{eq11}
F = \bar {h} .
\end{equation}

\noindent
where $\bar {h}$ is a constant of integration ,i.e., the total energy of the
system.

After the above treatment, finally we have the Hamiltonian of a one degree
of freedom system:

\begin{equation}
\label{eq12}
 F = \tilde {F}_{0} \left( {\tilde {L}} \right) + \tilde {F}_{c} \left(
{\tilde {L} , \tilde {G}_{0}} \right) + \tilde {F}_{1} \cos\tilde {l} ,
\end{equation}

\noindent
and we have the canonical Hamiltonian equations of motion,

\begin{equation}
\label{eq13}
\left\{
\begin{array}{l}
\dot{\tilde{L}}  =  \frac{{\partial F}}{{\partial \tilde {l}}} = - \tilde {F}_{1} \sin\tilde {l} \\
\dot {\tilde {l}} = - \frac{{\partial F}}{{\partial \tilde {L}}} =
- \left( {\frac{{\partial \tilde {F}_{0}} }{{\partial \tilde {L}}}
+ \frac{{\partial \tilde {F}_{c}} }{{\partial \tilde {L}}}
+ \frac{{\partial \tilde {F}_{1}} }{{\partial \tilde {L}}}\cos\tilde {l}} \right) . \\
\end{array}
\right.
\end{equation}

According to Eq. 13, we theoretically investigated the motion of the inner planet
using the initial values given in Table 1. Fig. 6 illustrates the
results of this analytical model. The resonant argument $\sigma $ librates
with an amplitude of $ \pm 70^{ \circ} $, and from the figure we are able to
estimate the equilibrium point to be about (0.131, 0). Moreover, Fig.6 also
indicates that the inner planet is really in the 2:1 resonant state with the
outer planet. Hence, in comparison with Fig.4, we come to the conclusion
that the structure of phase space given by our analytical model is
consistent with the leading results of our numerical simulations(see Fig.
4), and this model can help explaining the resonant mechanism of GJ 876,
which can maintain the stability of the system.

\subsection{The region of motion of the inner planet }
From the numerical simulations,,we have learned that the stability
of the system greatly depends on the initial configurations,
however in the perspective of the analytics, we still hope to get
knowledge about the dynamical evolution of the system as we change
other parameters by means of the proposed resonant model.

Rivera \& Lissauer(2001) provided the fitting results of the
planetary masses expressed by the product $M_{p}\sin i$ according
to their dynamical models , but in a sense the actual masses are
really hard to determine, so at first we aim to understand the
dynamics of the system if we gradually change the mass of the
outer planet $\mu _{2} $. In accordance with Eq. 13, we set to
carry out calculations to study the region of motion of the inner
planet by varying $\mu _{2} $. In our experiments, we let $\mu
_{2} $ =0.05,0.15,0.30,0.32,0.332,0.35,0.55,1.06,1.695 and 3.39
$M_{J} $ separately and other parameters are still chosen from
Table 1. Fig. 7 exhibits the portrait of the phase space of the
computations, then the semimajor axis $a$ is plotted against the
resonant argument $\sigma $ for different masses. From the phase
diagram, we note that when $\mu _{2} \le 0.32 M_{J} $, we have the
circulation of the system; for $3.39 \le \mu _{2} \le 0.332 M_{J}
$, in contrast, the system undergoes the state of libration
instead. The figure indicates that the perturbation of the outer
planet may greatly influence on the regime of motion of the inner
planet and the resonance becomes much stronger as the mass of the
outer planet $\mu _{2} $ increases, so that the region of motion
of the inner planet is much wider. Furthermore, we shall seek the
quantitative explanation. In fact, in terms of Eq. 13,if we fix
$e$,$a_{0}$ and $\tilde {l}_{0} $ and drive $\mu _{2} $ augment,
then  $\dot {\tilde {L}}$ increases, but $\dot {\tilde {l}}$
decreases, vice versa .

On the other hand, we studied the system by varying the starting
angle variable $\tilde {l}_{0} $ in  Eq. 7, which corresponds to
different cases of the initial motion of the two planets. For
detailed calculations,we took $\tilde {l}_{0} $ as 0$^{0}$,
30$^{0}$ ,60$^{0} $,90$^{0}$,
120$^{0}$,135$^{0}$,144$^{0}$,145$^{0} $,150$^{0}$ and
180$^{0}$,respectively. The dominant results are presented by Fig.
8. The figure displays that the intensity of the resonance grows
less weaker as the angle becomes much larger, which the dynamical
behavior of  the system varies from libration to circulation ;
especially for $\tilde {l}_{0} $=145$^{0}$, the trajectory lies on
the side of the separatrix, on the edge of the resonance. In
addition, for smaller $\tilde {l}_{0}$, we can easily detect that
the equilibrium point turns to be (0.131, 0),which again confirms
those of the numerical simulations. In a word, these quantitative
study may somewhat sketch rough outline of the dynamics of the GJ
876 system.

\section{Discussion}
In this paper, we have mainly explored the dynamics of the GJ 876 system by
simulating different cases of the planetary configurations. Moreover, we
also proposed an analytical model to make quantitative explanation of the
studied system.

In the end, we summarize some conclusions:

For two-planet coplanar systems, we found that nearly 1/4 of  hundreds of
systems in our simulations remain stable over a time span of 1Myr. Moreover,
the stability of a system is sensitive to its initial planetary
configuration. And the numerical results suggested the existence of a 2:1
resonance for the stable systems, and this kind of resonant mechanism can
also be found for those stable cases of the two-planet highly inclined orbit
systems. Therefore, according to the resulting simulations, we should emphasize
that the 2:1 mean motion resonance between two planets can act as an
effective mechanism for maintaining the stability of the system.

 From another direction, we have proposed an analytical model to explain the
resonant mechanism found in the simulations and observations. In a
qualitative viewpoint, the model again demonstrates that the 2:1 mean motion
resonance plays the major role of sustaining the system. Using the model, we
then studied the region of  motion of the inner planet by changing
different parameters involved in the dynamics of the system and presented
the helpful results.

However, the actual dynamics of the GJ 876 system should be rather
complicated beyond imagination. As mentioned before, the secular resonance
might also have effect on the long-term dynamical evolution of the system,
thus we can safely reach the conclusion that in a sense the mixture of the
different resonances are responsible for the dynamics of the system.
As we still do not know much about such systems,
we shall make further study of the dynamics of the extrasolar planets
to better understand their origin and evolution.

\acknowledgments

{We would thank Dr. John Chambers for carefully reading the
manuscript and for constructive remarks and suggestions.This work
is financially supported by the National Natural Science
Foundations of China(Grant No.10173006),the Research Fund for the
Doctoral Program(RFDP) of Higher Education(Grant No.2000028416),
the Foundation of the Postdoctoral Science of China and the
Foundation of  Minor Planets of Purple Mountain Observatory.}

\clearpage

\begin{center}
Figure Captions
\end{center}

Fig.1   The initial configuration of  two companions projected on the plane,
the red filled circle represents the main star; the blue, the inner planet;
the magenta, the outer planet. The orbital periods of are 30.1d and
61.0d,respectively.

\bigskip

Fig.2  The variations of the semimajor axis $a$, eccentricity and
inclination of the inner and outer planets for the time span of
1Myr.  The unit of the semimajor axis is AU, and degree for the
inclination. The  figure shows that the orbital parameters do not
change dramatically, and the semi-major axes $a$ librate about
0.130 AU and 0.210 AU,respectively, then the mean motion resonance
between two planets maintains during the dynamical evolution.The
evolution shown in the figure is typical of all the stable cases.

\bigskip

Fig.3  The critical argument $\sigma $ is plotted against  time, note
that the argument librates with an amplitude of $ \pm 70^{ \circ} $ over the
time span of 1Myr.

\bigskip

Fig.4  The motion of the phase space of the inner planet is given by numerical
simulations, the semimajor axis is plotted as a function of resonant
argument $\sigma $, note that the equilibrium point is around (0.130, 0).

\bigskip

Fig.5  The case of the secular resonance of the inner planet, the
upper panel exhibits that the eccentricity changes with time and
the lower panel shows that the periastron $\omega$ as a function
of time,temporarily librating about 0$^{0}$ or 180$^{0}$. In final
, the inner planet escapes from its orbit because of the
excitation of the eccentricity.

\bigskip

Fig.6  The motion of the phase space of the inner planet is presented by analytical method
, the semimajor axis is plotted as a function of resonant argument $\sigma $
,which also librates with an amplitude of $ \pm 70^{ \circ} $. In comparison
with Fig.4, the structure of phase space is in good agreement with each  other.

\bigskip

Fig.7  The motion of the phase space of the inner planet for different
masses of the outer planet, for $\mu _{2} \le 0.32 M_{J} $,the system
in the circulation; for $3.39 \le \mu _{2} \le 0.332 M_{J}$,
the system in the libration, the resonance becomes much stronger as
we  increase  the mass of the outer planet, so that the region of motion
of the inner planet is much wider .

\bigskip

Fig.8  The motion of the phase space of the inner planet for different
initial phases of  two planets, note that for $\tilde {l}_{0} $=145$^{0}$,
the trajectory lies on the side of the separatrix, on the edge of the resonance.
For smaller $\tilde {l}_{0}$, notice that the equilibrium point turns to be (0.131, 0).
\clearpage

\begin{deluxetable}{ccc}
\tablewidth{0pt}
\tablecaption{The inital osculating orbital elements of two planets}
\tablehead{
\colhead{Parameter}    & \colhead{Inner}      & \colhead{Outer}
}
\startdata
Mass($M_{Jup}$)  & 1.06 & 3.39 \\
Periods(days)  & 29.995 & 62.092 \\
$a$(AU) & 0.1294 & 0.2108 \\
Eccentricity & 0.314 & 0.051 \\
Inclination(deg) & 0.5 & 0.5 \\
$\Omega$(deg) & 59.2 & 20.0 \\
$\omega$(deg) & 51.8 & 40.0 \\
Mean anomaly(deg) & 289.0 & 340.0 \\
\enddata
\end{deluxetable}

\clearpage

\begin{deluxetable}{cccccc}
\tablewidth{0pt}
\tablecaption{The orbital elements variation of two planets}
\tablehead{
\colhead{Inner}    & \colhead{Upper limit}      & \colhead{Lower limit} &
\colhead{Outer}    & \colhead{Upper limit}      & \colhead{Lower limit}
}
\startdata
$a$(AU)  & 0.1353 & 0.1248 & $a$(AU) & 0.2148 & 0.2013  \\
$e$  & 0.3676 & 0.2419 & $e$ & 0.0631 & 0.0001  \\
$i(^\circ)$  & 0.7852 & 0.1806 & $i(^\circ)$ & 0.5560 & 0.4097  \\
\enddata
\end{deluxetable}

\clearpage

\begin{deluxetable}{lll}
\tablewidth{0pt}
\tablecaption{The status of GJ 876 system for different inclinations of the outer
planet over the time span of 1Myr}
\tablehead{
\colhead{Incl. of inner planet}  & \colhead{Incl. of outer planet}      &
\colhead{status of the system} }
\startdata
0.5  & 5.5 & Stable, MMR \\
\nodata  & 10.5 & T $>$ 413,560 yr , inner planet escape \\
\nodata  & 15.5 & T $>$ 158,880 yr , inner planet escape \\
\nodata  & 20.5 & T $>$ 553,172 yr , inner planet escape \\
\nodata  & 25.5 & Stable, MMR \\
\nodata  & 30.5 & Stable, MMR\\
\enddata
\end{deluxetable}


\begin{thebibliography}{}
\bibitem[Butler et al. (2000)]{But00}Butler, R. P., Vogt, S. S., Marcy,
G. W., Fischer, D. A., Henry, G. W., \& Apps, K. 2000, \apj, 545, 504

\bibitem[Duncan & Levison (1997)]{Dun97}Duncan, M. J., \& Levison H. F. 1997,
Science, 276, 1670

\bibitem[Feng(1986)]{Fen86}
Feng  K. 1986, J. Comp. Math.,4, 279

\bibitem[Fehlberg(1968)]{Feh68}
Fehlberg  E. 1968, NASA  TR  R-287


\bibitem[Garfinkel(1966)]{Gar66}Garfinkel, B. 1966, \aj, 71, 657
\bibitem[Ji et al.(2000)]{Ji00}Ji J. H., Liu. L. \& Liao X. H. 2000, Chin.
Astron. Astrophy., 24(4),399
\bibitem[Kozai(1966)]{Koz66}Kozai Y. 1962, \aj, 67, 591
\bibitem[Laughlin & Chambers(2001)]{Lau01}Laughlin, G. \& Chambers ,J. 2001,
\apj,551,L109

\bibitem[Liu et al.(1985a)]{Liu85a}Liu L. , Innanen, K. \& Zhang, S. P. 1985a,
\aj, 90 , 877

\bibitem[Liu et al.(1985b)]{Liu85b}Liu L. , Innanen, K. 1985b, \aj, 90 , 887
\bibitem[Malhotra(1996)]{Mal96}Malhotra, R. 1996 , \aj, 111, 504
\bibitem[Malhotra(1998)]{Mal98}
Malhotra, R. 1998, in ASP Conf. Proc. 149, Solar System Formation and
Evolution, ed. Lazzaro,et al.(San Frnacisco: ASP), 37

\bibitem[Malhotra et al.(2000)]{Mal00}
Malhotra, R., Duncan, M. J., \& Levison, H. 2000, in Protostars and Planets
IV, ed. V. Mannings, A. P. Boss \& S. S. Russell (Tucson: Univ. Arizona
Press), 1231

\bibitem[Marcy et al.(2001)]{Mar01}
Marcy, G. W., Butler, R. P., Fischer, D., Vogt, S. S., Lissauer, J. J., \&
Rivera, E. J. 2001, \apj, 556, 296

\bibitem[Marcy et al.(2000)]{Mar00}
Marcy, G.W. Cochran, W. D. \& Mayor, M. 2000, in Protostars and Planets IV,
ed. V. Mannings, A.P. Boss \& S. S. Russell (Tucson: University of Arizona
Press), p.1285

\bibitem[Morbidelli & Nesvorny(1999)]{Mor99}
Morbidelli, A. \& Nesvorny , D. 1999, Icarus , 139, 295

\bibitem[Rivera & Lissauer(2000)]{Riv00}
Rivera,E.J., \& Lissauer, J.J 2000, \apj, 530, 544

\bibitem[Rivera & Lissauer(2001)]{Riv01}
Rivera,E.J., \& Lissauer, J.J 2001, \apj, 558, 392

\bibitem[Wisdom & Holman(1991)]{Wis91}

Wisdom, J.,\& Holman, M. 1991,\aj,102,1528
\end{thebibliography}
\end{document}